\documentclass[runningheads]{llncs}
\usepackage{graphicx}
\usepackage{tikz}
\usepackage{pgfplots}
\usepackage{amsmath}
\usepackage{amsfonts}
\usepackage{mydefs}	
\usepackage{latexsym}

\begin{document}
\title{RevPHiSeg: A Memory-Efficient Neural Network for Uncertainty Quantification in Medical Image Segmentation}
%
%
\author{Marc Gantenbein \and
Ertunc Erdil \and
Ender Konukoglu}
\authorrunning{M.Gantenbein et al.}
\titlerunning{Reversible PHiSeg}
%
\institute{Computer Vision Laboratory, ETH Zürich, Switzerland
\email{\{ertunc.erdil,ender.konukoglu\}@vision.ee.ethz.ch}}
\maketitle              

\begin{abstract}
Quantifying segmentation uncertainty has become an important issue in medical image analysis due to the inherent ambiguity of anatomical structures and its pathologies. Recently, neural network-based uncertainty quantification methods have been successfully applied to various problems. One of the main limitations of the existing techniques is the high memory requirement during training; which limits their application to processing smaller field-of-views (FOVs) and/or using shallower architectures. In this paper, we investigate the effect of using reversible blocks for building memory-efficient neural network architectures for quantification of segmentation uncertainty. The reversible architecture achieves memory saving by exactly computing the activations from the outputs of the subsequent layers during backpropagation instead of storing the activations for each layer. We incorporate the reversible blocks into a recently proposed architecture called PHiSeg that is developed for uncertainty quantification in medical image segmentation. The reversible architecture, RevPHiSeg, allows training neural networks for quantifying segmentation uncertainty on GPUs with limited memory and processing larger FOVs. We perform experiments on the LIDC-IDRI dataset and an in-house prostate dataset, and present comparisons with PHiSeg. The results demonstrate that RevPHiSeg consumes $\sim30\%$ less memory compared to PHiSeg while achieving very similar segmentation accuracy.

\end{abstract}

\section{Introduction}

Segmentation has been a crucial problem in medical image analysis for clinical diagnosis and many downstream tasks. The majority of the segmentation algorithms in the literature aim to find a single segmentation as a solution which is a point estimate in the posterior distribution of a segmentation given an image \cite{ronneberger2015unet}. However, having a point estimate does not provide a measure of the degree of confidence in that result, neither does it provide a picture of other probable solutions based on the data and the priors. Due to the inherent ambiguities and uncertainties in many medical images, characterization of the posterior distribution through its samples plays a crucial role in quantifying the uncertainty and revealing other plausible segmentations.

There have been some efforts in the literature for generating segmentation samples from the underlying posterior distribution. One group of methods aims at using Markov chain Monte Carlo (MCMC) techniques which ensure asymptotic convergence to the desired posterior \cite{erdil2016mcmc,fan2007mcmc,erdil2019pseudomcmc}. However, these methods suffer from slow convergence and satisfying the necessary conditions of MCMC is non-trivial \cite{fan2007mcmc,erdil2016mcmc}. Another group of methods is based on variational inference which approximate the desired posterior density using a variational function. One of the pioneering variational inference-based methods that shows significant performance for segmentation uncertainty quantification is Probabilistic U-Net by Kohl et al. \cite{kohl2018probabilistic}. The method minimizes the Kullback-Leibler (KL) divergence between a posterior and a prior network during training. Then, the samples are generated from the learned latent distribution and appended to the penultimate layer of a U-Net \cite{ronneberger2015unet} to generate segmentation samples. Recently, Baumgartner et al. \cite{PHiSeg2019Baumgartner} proposed a method called PHiSeg that samples from the learned distributions in every latent layer of a U-Net instead of the last latent level as in Probabilistic U-Net. PHiSeg achieves better performance compared to Probabilistic U-Net on various medical image segmentation datasets.

Although, both Probabilistic U-Net and PHiSeg achieve promising performance in terms of segmentation quality and uncertainty quantification, they suffer from a significant memory burden during training. This either limits their domain of applicability to processing images with small field-of-view which is not desired especially in medical domain or requires GPUs with large memories which are very expensive to obtain. To overcome this limitation, we investigate using reversible blocks for building a memory efficient architecture that generates segmentation samples for uncertainty quantification in medical image segmentation. To achieve this, we incorporate reversible blocks \cite{revresnet} into PHiSeg for a smaller memory consumption during training and built a new architecture called RevPHiSeg. Reversible blocks have been previously used along with U-Net \cite{ronneberger2015unet} for segmentation \cite{PartiallyRevUnet2019Bruegger}. They allow us to recover the exact activation of each layer from the following layer during training and eliminate the need to store activations for each layer during backpropagation.

We perform experiments on two different datasets: LIDC-IDRI \cite{lidc-data,lidcidri2011} and an in-house prostate data set. The results demonstrate that RevPHiSeg achieves $\sim30\%$ less memory consumption compared to PHiSeg by achieving very competitive results in terms of the quality of segmentation samples. The implementation of RevPHiSeg will be made available.\footnote{\url{https://github.com/gigantenbein/UNet-Zoo}}

\section{Methods}
In this section, we present RevPHiSeg after providing some background information on PHiSeg \cite{PHiSeg2019Baumgartner} and reversible blocks \cite{revresnet} for the sake of completeness.

\subsection{PHiSeg}
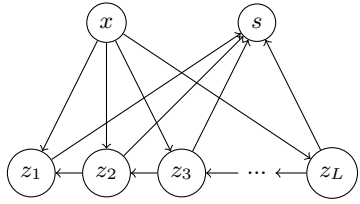
\begin{figure}
    \centering
    \begin{tikzpicture}
            \node (z1) at (0,0) [circle,draw]{$z_1$};
            \node (z2) at (1,0) [circle,draw]{$z_2$};
            \node (z3) at (2,0) [circle,draw]{$z_3$};
            \node (zL) at (4,0) [circle,draw]{$z_L$};
            \node (dots) at (3,0) {$...$};
            \node (x) at (1,2) [circle,draw] {$x$};
            \node (s) at (3,2) [circle,draw] {$s$};
            
            \draw[-{To}] (x) -- (z1);
            \draw[-{To}] (x) -- (z2);
            \draw[-{To}] (x) -- (z3);
            \draw[-{To}] (x) -- (zL);
            
            \draw[-{To}] (z1)  -- (s);
            \draw[-{To}] (z2) -- (s);
            \draw[-{To}] (z3) -- (s);
            \draw[-{To}] (zL) -- (s);
            
            \draw[-{To}] (zL) -- (dots);
            \draw[-{To}] (dots) -- (z3);
            \draw[-{To}] (z3) -- (z2);
            \draw[-{To}] (z2) -- (z1);
        \end{tikzpicture}
    \caption{Graphical model for hierarchical segmentation}
    \label{fig:hierarchical_model}
\end{figure}

PHiSeg aims to approximate the posterior distribution $p(\mathbf{z}|\mathbf{s},\mathbf{x})$ using a variational function as in \cite{kingma2013autoencoding}, where $\mathbf{x}$ is the input image and $\mathbf{s}$ is the segmentation, and $\mathbf{z}$ is the latent representation. In PHiSeg, the latent variable $\mathbf{z} = \lbrace \mathbf{z}_1,\dotso, \mathbf{z}_L \rbrace$ is modeled hierarchically as shown in the graphical model in Fig. \ref{fig:hierarchical_model}. 

Then, the posterior distribution of the segmentation $\mathbf{s}$ given an image $\mathbf{x}$ can be written for the general case of $L$ latent levels as:
\begin{equation}{
p(\mathbf{s}|\mathbf{x}) = \int\limits_{\mathbf{z}_1,\dotso, \mathbf{z}_L} p(\mathbf{s}|\mathbf{z}_1,\dotso,\mathbf{z}_L)p(\mathbf{z}_1|\mathbf{z}_2,\mathbf{x})\dotsm p(\mathbf{z}_{\text{L-1}}|\mathbf{z}_L, \mathbf{x})p(\mathbf{z}_L|\mathbf{x}) d\mathbf{z}_1\dotso d\mathbf{z}_L
}
\end{equation}
The posterior distribution $p(\mathbf{z}|\mathbf{s}, \mathbf{x})$ can be approximated by a variational function $q(\mathbf{z}|\mathbf{s},\mathbf{x})$ using variational inference. Minimizing the Kullback-Leibler ($\KL$) divergence between $p(\mathbf{z}|\mathbf{s}, \mathbf{x})$ and $q(\mathbf{z}|\mathbf{s}, \mathbf{x})$ results in the following lower-bound estimate of $\log p(\mathbf{s}|\mathbf{x})$: 
\begin{equation}\label{eq:loss}{
    \log{p(\mathbf{s}|\mathbf{x}}) = \mathcal{L}(\mathbf{s}|\mathbf{x}) + \KL(q(\mathbf{z}|\mathbf{s},\mathbf{x})||p(\mathbf{z}|\mathbf{s},\mathbf{x}))
}
\end{equation}
where, $\mathcal{L}$ is a lower-bound on $\log p(\mathbf{s}|\mathbf{x})$ with equality when the approximation $q$ matches the posterior exactly. The lower bound $\mathcal{L}(\mathbf{s}|\mathbf{x})$ can be written as
\begin{equation}
    \begin{split}
        \mathcal{L}(\mathbf{s}|\mathbf{x}) &= \mathbb{E}_{q(\mathbf{z}_1,\dotso, \mathbf{z}_L|\mathbf{x},\mathbf{s})}[\log{p(\mathbf{s}|\mathbf{z}_1,\dotso,\mathbf{z}_L)}] - \alpha _L \KL(q(\mathbf{z}_L|\mathbf{s},\mathbf{x})||p(\mathbf{z}_L|\mathbf{x})) \\
        & -\sum_ {l=1}^{L-1} \alpha_l \mathbb{E}_{q(\mathbf{z}_{l+1}|\mathbf{s},\mathbf{x})}
        [\KL[q(\mathbf{z}_l|\mathbf{z}_{l+1},\mathbf{s},\mathbf{x})||p(\mathbf{z}_l|\mathbf{z}_{l+1},\mathbf{x})]]
    \end{split}
\end{equation}
where, $\alpha_i$ is a weighting term which we set to 1 in our experiments.

In PHiSeg \cite{PHiSeg2019Baumgartner} the distributions $p(\mathbf{z}_l|\mathbf{z}_{L-1},\mathbf{x})$ and $q(\mathbf{z}_l|\mathbf{z}_{L-1},\mathbf{x},\mathbf{s})$ are parametrized by axis-aligned normal distributions, which are defined as follows:
\begin{equation}
    p(\mathbf{z}_l|\mathbf{z}_{l+1},\mathbf{x}) = \mathcal{N}(\mathbf{z}|\phi_l^{(\mu)}(\mathbf{z}_{l+1},\mathbf{x}),\phi_l^{(\sigma)}(\mathbf{z}_{l+1},\mathbf{x}))
\end{equation}
\begin{equation}
    q(\mathbf{z}_l|\mathbf{z}_{l+1},\mathbf{x}, \mathbf{s}) = \mathcal{N}(\mathbf{z}|\theta_l^{(\mu)}(\mathbf{z}_{l+1},\mathbf{x}, \mathbf{s}),\theta_l^{(\sigma)}(\mathbf{z}_{l+1},\mathbf{x}, \mathbf{s}))
\end{equation}
where the functions $\phi$ and $\theta$ are parametrized by neural networks. The  architecture is trained by minimizing Eq.\ref{eq:loss}. 

\subsection{Reversible architectures}
\label{sec:reversible_architectures}
One of the major reasons for memory consumption in neural networks is due to storing the activations during the forward pass to be used in backpropagation. To alleviate the memory burden of the stored activations, Gomes et al. \cite{revresnet} propose reversible blocks that allow achieving memory savings by avoiding to store the activations in the forward pass by using reversible layers. Instead, the activations are computed from the previous layers when needed during backpropagation.

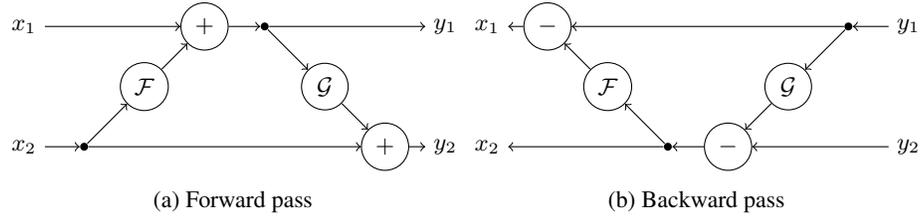
\begin{figure}
    \begin{subfigure}[b]{0.5\textwidth}
            \begin{tikzpicture}[scale=0.8]
                \node (x1) at (0,2) {$x_1$};
                \node (x2) at (0,0) {$x_2$};
                \node (p1) [circle,draw=black]at (3,2) {$+$};
                \node (c1) [circle,fill=black,inner sep=0pt,minimum size=3pt] at (4,2){};
                
                \node (f) [circle,draw=black]at (2,1) {$\mathcal{F}$};
                \node (g) [circle,draw=black]at (5,1) {$\mathcal{G}$};
                \node (p2) [circle,draw=black]at (6,0) {$+$};
                \node (c2) [circle,fill=black,inner sep=0pt,minimum size=3pt] at (1,0){};
                \node (y1) at (7,2) {$y_1$};
                \node (y2) at (7,0) {$y_2$};
                
                \draw[-{To}] (x1) -- node[auto] {} (p1);
                \draw[-{To}] (p1) -- node[auto] {} (c1);
                \draw[-{To}] (c1) -- node[auto] {} (y1);
                \draw[-{To}] (c1) -- node[auto] {} (g);
                \draw[-{To}] (g) -- node[auto] {} (p2);
                
                \draw[-{To}] (x2) -- node[auto] {} (c2);
                \draw[-{To}] (c2) -- node[auto] {} (p2);
                \draw[-{To}] (c2) -- node[auto] {} (f) ;
                \draw[-{To}] (f) -- node[auto] {} (p1) ;
                \draw[-{To}] (p2) -- node[auto] {} (y2);
            \end{tikzpicture}
        \caption{Forward pass}
        \label{fig:forward_rev}
    \end{subfigure}
    \begin{subfigure}[b]{0.5\textwidth}
        \begin{tikzpicture}[scale=0.8]
            \node (x1) at (0,2) {$x_1$};
            \node (x2) at (0,0) {$x_2$};
            \node (p1) [circle,draw=black]at (1,2) {$-$};
            \node (c1) [circle,fill=black,inner sep=0pt,minimum size=3pt] at (6,2){};
            
            \node (f) [circle,draw=black]at (2,1) {$\mathcal{F}$};
            \node (g) [circle,draw=black]at (5,1) {$\mathcal{G}$};
            \node (p2) [circle,draw=black]at (4,0) {$-$};
            \node (c2) [circle,fill=black,inner sep=0pt,minimum size=3pt] at (3,0){};
            \node (y1) at (7,2) {$y_1$};
            \node (y2) at (7,0) {$y_2$};
            
            \draw[-{To}] (y1) -- node[auto] {} (c1);
            \draw[-{To}] (c1) -- node[auto] {} (g);
            \draw[-{To}] (c1) -- node[auto] {} (p1);
            \draw[-{To}] (g) -- node[auto] {} (p2);
            \draw[-{To}] (p1) -- node[auto] {} (x1) ;
            
            \draw[-{To}] (f) -- node[auto] {} (p1);
            \draw[-{To}] (c2) -- node[auto] {} (f);
            \draw[-{To}] (c2) -- node[auto] {} (x2) ;
            \draw[-{To}] (p2) -- node[auto] {} (c2) ;
            \draw[-{To}] (y2) -- node[auto] {} (p2);
        \end{tikzpicture}
        \caption{Backward pass}
        \label{fig:backwar_rev}
    \end{subfigure}
    \caption{Sketch of the computations of forward and backward passes in reversible blocks.}
    \label{fig:rev_blocks}
\end{figure}

A reversible block consists of two functions $\mathcal{F}$ and $\mathcal{G}$ as shown in Fig. \ref{fig:rev_blocks}. $\mathcal{F}$ and $\mathcal{G}$ can be arbitrary functions such as a fully-connected layer, a convolutional layer or a nonlinearity. The functions are expected to have the same input and output dimensions since the performed operations have to be invertible.

Fig. \ref{fig:rev_blocks} shows the computations which take place during the forward (in Fig. \ref{fig:forward_rev}) and the backward (in Fig.\ref{fig:backwar_rev}) passes. Gomes et al. \cite{revresnet} partitions the input of the reversible block into two groups. The first group $x_1$ flows from the top batch while the second group uses the bottom one as shown in Fig. \ref{fig:rev_blocks}. The inputs $x_1$ and $x_2$ are recovered after the backward pass by inverting the operations in the forward pass with the gradients of $\mathcal{F}$ and $\mathcal{G}$ with respect to their parameters.

\subsection{RevPHiSeg}

To create a memory-efficient architecture for quantification of segmentaion uncertainty, we
incorporated reversible blocks into PHiSeg. When looking for an option to employ reversible blocks, the convolutional layer offer themselves due to the number of activations stored in them. To achieve memory savings, multiple reversible blocks have to be concatenated which then form a reversible sequence.
Considering Fig.\ref{fig:rev_blocks}, the functions $\mathcal{F}$ and $\mathcal{G}$ would then correspond to a $3 \times 3$ convolutional layers with a nonlinearity. For our architecture RevPHiSeg, each sequence of convolutions in the original PHiSeg was replaced by a sequence of reversible blocks.

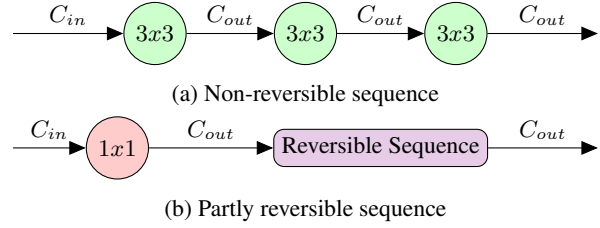
\begin{figure}[h]
    \begin{subfigure}{\textwidth}
    \centering
            \begin{tikzpicture}
                \node (in) at (0,0) {};
                \node (c1) [circle,draw=black, fill=green!20]at (2,0) {$3x3$};
                \node (c2) [circle,draw=black, fill=green!20]at (4,0) {$3x3$};
                \node (c3) [circle,draw=black, fill=green!20]at (6,0) {$3x3$};
                \node (out) at (8,0) {};
                
                \draw[->] (in) -- node[auto] {$C_{in}$} (c1);
                \draw[->] (c3) -- node[auto] {$C_{out}$} (out);
                \draw[->] (c1) -- node[auto] {$C_{out}$} (c2) ;
                \draw[->] (c2) -- node[auto] {$C_{out}$} (c3);
            \end{tikzpicture}
        \caption{Non-reversible sequence}
        \label{fig:non_reversible_sequence}
    \end{subfigure}
    \begin{subfigure}{\textwidth}
        \centering
        \begin{tikzpicture}
            \node (in) at (0,0) {};
            \node (c1) [circle,draw=black, fill=red!20]at (1.5,0) {$1x1$};
            \node (c2) [rounded corners,draw=black, fill=violet!20]at (5,0) {$\text{Reversible Sequence}$};
            \node (out) at (8,0) {};
            
            \draw[->] (in) -- node[auto] {$C_{in}$} (c1);
            \draw[->] (c2) -- node[auto] {$C_{out}$} (out);
            \draw[->] (c1) -- node[auto] {$C_{out}$} (c2);
        \end{tikzpicture}
        \caption{Partly reversible sequence}
        \label{fig:reversible_sequence}
    \end{subfigure}
    \caption{Replacing a series of convolutions with a reversible sequence}
    \label{fig:comparison_reversible_sequence}
\end{figure}

As stated in Sec.\ref{sec:reversible_architectures}, replacing a function with a reversible
block requires the function to have the same input and out dimensions. For convolutions, this implies that the convolution needs the same number of input and output channels. However, PHiSeg resembles a UNet architecture and thus contains convolutions with a different number of input and output channels.
We thus decided to add a $1 \times 1$ convolution before every reversible sequence where the number of channels do change. The $1 \times 1$ convolution have different input and output dimensions and thus enable having the same input and output dimensions for the following $3 \times 3$ convolutions in the reversible blocks as can be seen in Fig.\ref{fig:comparison_reversible_sequence}.

\begin{figure}[h]
    \centering
    \includegraphics[width=\textwidth]{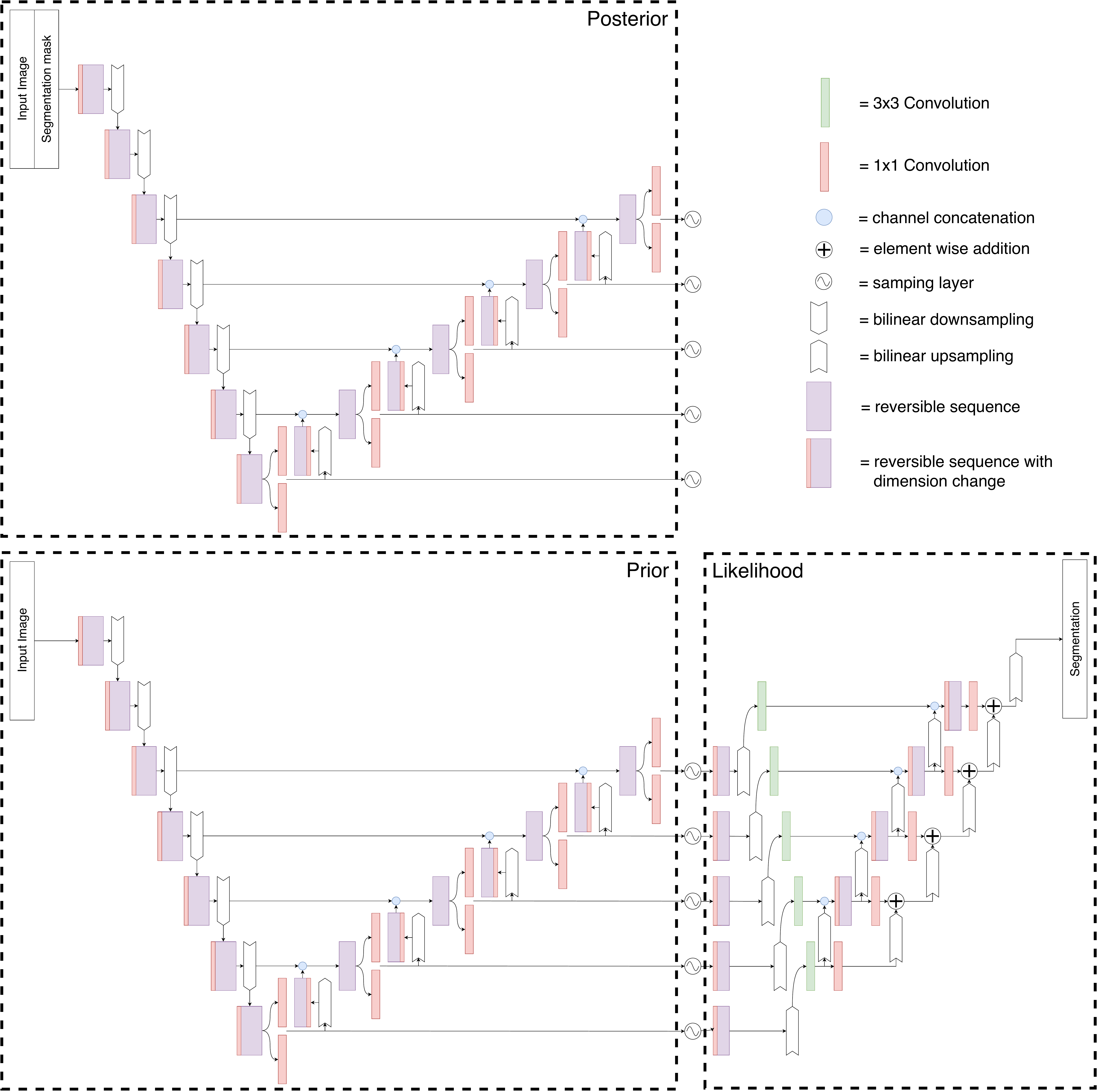}
    \caption{Revesible PHiSeg with 5 latent levels and 7 resolution levels with corresponding Posterior, Prior and Likelihood network for training and inference}
    \label{fig:phiseg_rev}
\end{figure}

Applying the same process to each convolutional sequence in PHiSeg, we obtain the RevPHiSeg architecture shown in Fig. \ref{fig:phiseg_rev}. As one can see, each convolutional sequence that features a dimension change is preceded by a $1 \times 1$ convolution.

\section{Experimental results}
In this section, we present the experimental results of RevPHiSeg on two different data sets: LIDC-IDRI \cite{lidc-data,lidcidri2011} and an in-house prostate dataset. We compare the results with PHiSeg in terms of both segmentation accuracy and memory consumption.

In our experiments, we chose the same number of latent and resolution levels as Baumgartner et al. to demonstrate the memory advantage of reversible blocks in PHiSeg architecture. Therefore, our configuration consisted of 5 latent layers, where a sampling step takes place, and 7 resolution layers with filters from 32 up to 192. 


\subsection{Evaluation metrics}

We obtain the quantitative results indicating the segmentation accuracy by comparing the ground truth segmentations of different experts with the segmentation samples generated by PHiSeg and RevPHiSeg. We exploit various evaluation metrics in our evaluations: Generalised Energy Distance (GED), Normalised Cross-Correlation (NCC) of the expected cross entropy between the mean segmentations and the mean of the generated samples, and the Dice score (DSC)  \cite{dice1945measures} per label.

GED is defined as 
\begin{equation}
    D_{GED}^2(p_{gt},p_s) =
    \frac{2}{nm} \sum_ {i=1}^{n} \sum_{j=1}^{m}  d(\mathbf{s}_i,\mathbf{y}_j)
    -\frac{1}{n^2}\sum_ {i=1}^{n} \sum_{j=1}^{n} d(\mathbf{s}_i,\mathbf{s}_j)
    -\frac{1}{m^2}\sum_ {i=1}^{m} \sum_{j=1}^{m} d(\mathbf{y}_i,\mathbf{y}_j)
\end{equation}
where, $d(\cdot,\cdot)=1-\IoU(\cdot,\cdot)$ with $\IoU(\cdot,\cdot)$ as the intersection over union, $m$ is the number of ground truth labels, $n$ is the number of segmentation samples, y are the ground truth samples and s are the segmentation samples. The generalized energy distance is a measure of the distance between two probability distribution where we treat the generated segmentations as samples from the approximate distribution and the ground truth labels as samples from the ground truth distribution.
\newline

To quantify the pixel-wise differences between the samples and ground truths, we use the normalized cross-correlation(NCC) of the cross entropy between the mean of the ground truth labels ($\bar{\mathbf{y}}$) and the mean of the generated samples ($\bar{\mathbf{s}}$), which is defined as follows:
\begin{equation}
    S_{NCC}(p_{gt},p_s)=\mathbb{E}_{\mathbf{y}\sim p_{gt}}[\NCC(\mathbb{E}_{\mathbf{s} \sim p_s}[\CE(\bar{\mathbf{s}},\mathbf{s})], \mathbb{E}_{\mathbf{s} \sim p_s}[\CE(\bar{\mathbf{y}},\mathbf{s})])
\end{equation}
where $p_{gt}$ is the ground truth distribution and $p_s$ is the approximate distribution \cite{PHiSeg2019Baumgartner}.

Furthermore, we use the Dice score (DSC) to measure the segmentation accuracy of each sample. DSC is a common metric that is used to quantify segmentation accuracy based on the overlap between the ground truth and the segmentation. When computing DSC we randomly choose a ground truth segmentation among different expert annotations, and calculate the DSC between the selected ground truth and the segmentation samples. We then average the DSC of these multiple draws.

We measure the memory consumption of the methods using the PyTorch function \texttt{max\_memory\_allocated()}. PHiSeg is originally implemented in Tensorflow. To have a fair comparison, we re-implemented PHiSeg in PyTorch.

\subsection{Datasets}
The LIDC-IDRI dataset contains 1018 lung CT scans each annotated by 4 radiologists. We use the same preprocessing as in \cite{kohl2018probabilistic,PHiSeg2019Baumgartner} and crop a $128 \times 128$ squares centered around the lesions. 

The prostate data set is an in-house data set that contains MR images from 68 patients. Each image in the data set has 6 annotations from 4 radiologists and 2 non-radiologists. We processed the data slice-by-slice (approx. 25 slices per volume), where we resampled each slice to a resolution of $0.6 \times 0.6 mm^2$ and took a central crop of size $192\times192$. 

We divide both datasets into a training, testing and validation set using a random 60-20-20 split.

\subsection{Experimental setup}
In our experiments, we use the network architecture shown in Fig. \ref{fig:phiseg_rev} for PHiSeg where we use 5 latent levels, $L = 5$, to generate samples as proposed for PHiSeg in \cite{PHiSeg2019Baumgartner}. The architecture we use for PHiSeg is similar to RevPHiSeg except for the reversible sequences and $1 \times 1$ convolutions. 

We use the Adam optimizer with a learning rate of $10^{-3}$. Furthermore, we used a weight decay of $10^{-5}$. We use ReLU and batch normalization after each convolutional layer on non-output layer. We train both PHiSeg and RevPHiSeg for 48 hours on an NVIDIA Titan X Pascal. While models are being trained, we calculate GED, NCC, and DSC on the validation sets. After 48h of training is done, we choose the model with the lowest GED score on the validation set. Finally, we evaluate the selected model on test sets to obtain the quantitative results. We conduct our experiments using an NVIDIA Titan Xp GPU with 12 GB of memory.

\subsection{Experimental results}
We present the quantitative results on LIDC-IDRI dataset in Tab.  \ref{tab:lidc_table}. The quantitative results demonstrate that RevPHiSeg achieves almost 30 percent memory savings while being quite competitive with PHiSeg in terms of the segmentation quality. 

Using the memory saving achieved by RevPHiSeg, we can process batch sizes of up to 56, where PHiSeg runs out of memory  after batch size 48. Although, we do not observe any improvement in terms of the segmentation quality when using larger batch sizes, being able to process larger batches could lead to improvement depending on the application such as unsupervised contrastive learning \cite{chen2020simple}.

\begin{table}
\centering
\caption{Quantitative results of RevPHiSeg and PHiSeg on LIDC-IDRI dataset.}
\begin{tabular}{l|l|lll|l|lll|l|}
                                &            & \multicolumn{4}{c|}{\textbf{LIDC-IDRI}} \\
                                & Batch size & $D_{GED}^2$     & $S_{NCC}$    & Dice    & Memory (MB)    \\ \hline
\multicolumn{1}{|l|}{PHiSeg}&\multicolumn{1}{c|}{12}&\multicolumn{1}{c}{0.2139}&\multicolumn{1}{c}{0.8533}&\multicolumn{1}{c|}{0.4991}&\multicolumn{1}{c|}{3251}      \\
\multicolumn{1}{|l|}{RevPHiSeg}&\multicolumn{1}{c|}{12}&\multicolumn{1}{c}{0.2365}&\multicolumn{1}{c}{0.7943}&\multicolumn{1}{c|}{0.5220}&\multicolumn{1}{c|}{\textbf{2194}}     \\ \hline
\multicolumn{1}{|l|}{PHiSeg}&\multicolumn{1}{c|}{24}&\multicolumn{1}{c}{0.2342}&\multicolumn{1}{c}{0.8296}&\multicolumn{1}{c|}{0.5344}&\multicolumn{1}{c|}{6076}      \\
\multicolumn{1}{|l|}{RevPHiSeg}&\multicolumn{1}{c|}{24}&\multicolumn{1}{c}{0.2396}&\multicolumn{1}{c}{0.7846}&\multicolumn{1}{c|}{0.5525}&\multicolumn{1}{c|}{\textbf{4070}}      \\ \hline
\multicolumn{1}{|l|}{PHiSeg}&\multicolumn{1}{c|}{36}&\multicolumn{1}{c}{0.2166}&\multicolumn{1}{c}{0.8387}&\multicolumn{1}{c|}{0.5229}&\multicolumn{1}{c|}{8905}      \\
\multicolumn{1}{|l|}{RevPHiSeg}&\multicolumn{1}{c|}{36}&\multicolumn{1}{c}{0.2677}&\multicolumn{1}{c}{0.7839}&\multicolumn{1}{c|}{0.4995}&\multicolumn{1}{c|}{\textbf{5903}}      \\ \hline
\multicolumn{1}{|l|}{PHiSeg}&\multicolumn{1}{c|}{48}&\multicolumn{1}{c}{0.2239}&\multicolumn{1}{c}{0.8409}&\multicolumn{1}{c|}{0.5224}&\multicolumn{1}{c|}{11374}     \\
\multicolumn{1}{|l|}{RevPHiSeg}&\multicolumn{1}{c|}{48}&\multicolumn{1}{c}{0.2436}&\multicolumn{1}{c}{0.8069}&\multicolumn{1}{c|}{0.5459}&\multicolumn{1}{c|}{\textbf{7948}}     \\ \hline
\multicolumn{1}{|l|}{PHiSeg}&\multicolumn{1}{c|}{56}&\multicolumn{1}{c}{-}&\multicolumn{1}{c}{-}&\multicolumn{1}{c|}{-}&\multicolumn{1}{c|}{-}    \\
\multicolumn{1}{|l|}{RevPHiSeg}&\multicolumn{1}{c|}{56}&\multicolumn{1}{c}{0.2478}&\multicolumn{1}{c}{0.7721}&\multicolumn{1}{c|}{0.5361}&\multicolumn{1}{c|}{\textbf{9238}}    \\
\hline
\end{tabular}
\label{tab:lidc_table}
\centering
\caption{Quantitative results of RevPHiSeg and PHiSeg on an in-house prostate dataset.}
\begin{tabular}{l|l|lll|l|lll|l|}
                                &            & \multicolumn{4}{c|}{\textbf{Prostate dataset}} \\
                                & Resolution &  $D_{GED}^2$     & $S_{NCC}$    & Dice    & Memory    \\ \hline
\multicolumn{1}{|l|}{PHiSeg}&\multicolumn{1}{c|}{192}&\multicolumn{1}{c}{0.3578}&\multicolumn{1}{c}{\textbf{0.7801}}&\multicolumn{1}{c|}{0.7480}&\multicolumn{1}{c|}{6813}      \\
\multicolumn{1}{|l|}{RevPHiSeg}&\multicolumn{1}{c|}{192}&\multicolumn{1}{c}{0.3035}&\multicolumn{1}{c}{0.75}&\multicolumn{1}{c|}{0.7871}&\multicolumn{1}{c|}{\textbf{4621}}     \\ \hline
\multicolumn{1}{|l|}{PHiSeg}&\multicolumn{1}{c|}{256}&\multicolumn{1}{c}{-}&\multicolumn{1}{c}{-}&\multicolumn{1}{c|}{-}&\multicolumn{1}{c|}{-}      \\
\multicolumn{1}{|l|}{RevPHiSeg}&\multicolumn{1}{c|}{256}&\multicolumn{1}{c}{\textbf{0.2486}}&\multicolumn{1}{c}{0.6712}&\multicolumn{1}{c|}{0.7094}&\multicolumn{1}{c|}{7582}      \\ \hline
\end{tabular}
\label{tab:uzh_table}
\end{table}

We present the quantitative results obtained on the in-house prostate dataset in Table \ref{tab:uzh_table}. The reversible architecture achieves significant memory saving compared to the vanilla PHiSeg.

The memory saving achieved by RevPHiSeg allows us to process images with higher resolutions. We perform experiments with two different resolutions: $192 \times 192$ and $256 \times 256$. While RevPHiSeg can process both resolutions, PHiSeg cannot process resolutions higher than $192 \times 192$. Although processing higher resolutions lead to a better score in terms of GED, the NCC and DSC results get slightly worse. This may be caused due to the architecture used for $192 \times 192$ is not large enough to learn effectively from $256 \times 256$ images or that the training time of 48h was not long enough for the resolution of $256 \times 256$.

\section{Discussion and Conclusion}
We investigate using reversible blocks for building a memory-efficient neural network architecture for generating segmentation samples to quantify segmentation uncertainty in medical images. To this end, we modified a state-of-the-art method, PHiSeg, by adding reversible blocks to make it memory efficient. We present quantitative results on two different medical datasets. The results demonstrate that RevPHiSeg consumes significantly less memory compared to the non-reversible architecture PHiSeg. The memory saving enables training RevPHiSeg on GPUs with limited memory, processing larger resolutions and using larger batches. Besides the memory saving, RevPHiSeg is quite competitive with PHiSeg in terms of segmentation quality. 



%
%
%
\bibliographystyle{splncs04}
\bibliography{references.bib}
\end{document}